\documentclass[10pt,conference]{IEEEtran}
\IEEEoverridecommandlockouts
\usepackage{cite}
\usepackage{amsmath,amssymb,amsfonts}
\usepackage{graphicx}
\usepackage{textcomp}
\usepackage{xcolor}
\usepackage{amsmath}
\usepackage{hyperref}
\usepackage{braket}
\usepackage{svg}

\usepackage[ruled, lined, commentsnumbered, longend]{algorithm2e}
\usepackage{algpseudocode}

\SetAlFnt{\mdseries\rmfamily}
\def\BibTeX{{\rm B\kern-.05em{\sc i\kern-.025em b}\kern-.08em
    T\kern-.1667em\lower.7ex\hbox{E}\kern-.125emX}}
\begin{document}

\title{Quantum Kernel Estimation With Neutral Atoms For Supervised Classification: A Gate-Based Approach\\

}

\author{\IEEEauthorblockN{Marco Russo, Edoardo Giusto, Bartolomeo Montrucchio}
\IEEEauthorblockA{\textit{Department of Computer and Control Engineering} \\
\textit{Politecnico di Torino}\\
Turin, Italy \\
\{marco.russo, edoardo.giusto, bartolomeo.montrucchio\}@polito.it}
\thanks{© 2023 IEEE.  Personal use of this material is permitted.  Permission from IEEE must be obtained for all other uses, in any current or future media, including reprinting/republishing this material for advertising or promotional purposes, creating new collective works, for resale or redistribution to servers or lists, or reuse of any copyrighted component of this work in other works.}
}

\maketitle

\begin{abstract}
Quantum Kernel Estimation (QKE) is a technique based on leveraging a quantum computer to estimate a kernel function that is classically difficult to calculate, which is then used by a classical computer for training a Support Vector Machine (SVM).
Given the high number of 2-local operators necessary for realizing a feature mapping hard to simulate classically, a high qubit connectivity is needed, which is not currently possible on superconducting devices.
For this reason, neutral atom quantum computers can be used, since they allow to arrange the atoms with more freedom.
Examples of neutral-atom-based QKE can be found in the literature, but they are focused on graph learning and use the analogue approach.
In this paper, a general method based on the gate model is presented.
After deriving 1-qubit and 2-qubit gates starting from laser pulses, a parameterized sequence for feature mapping on 3 qubits is realized.
This sequence is then used to empirically compute the kernel matrix starting from a dataset, which is finally used to train the SVM. It is also shown that this process can be generalized up to N qubits taking advantage of the more flexible arrangement of atoms that this technology allows.
The accuracy is shown to be high despite the small dataset and the low separation. 
This is the first paper that not only proposes an algorithm for explicitly deriving a universal set of gates but also presents a method of estimating quantum kernels on neutral atom devices for general problems using the gate model.

\end{abstract}


\section{Introduction}
\begin{figure}[t]

\centerline{\includegraphics[width=0.47\textwidth]{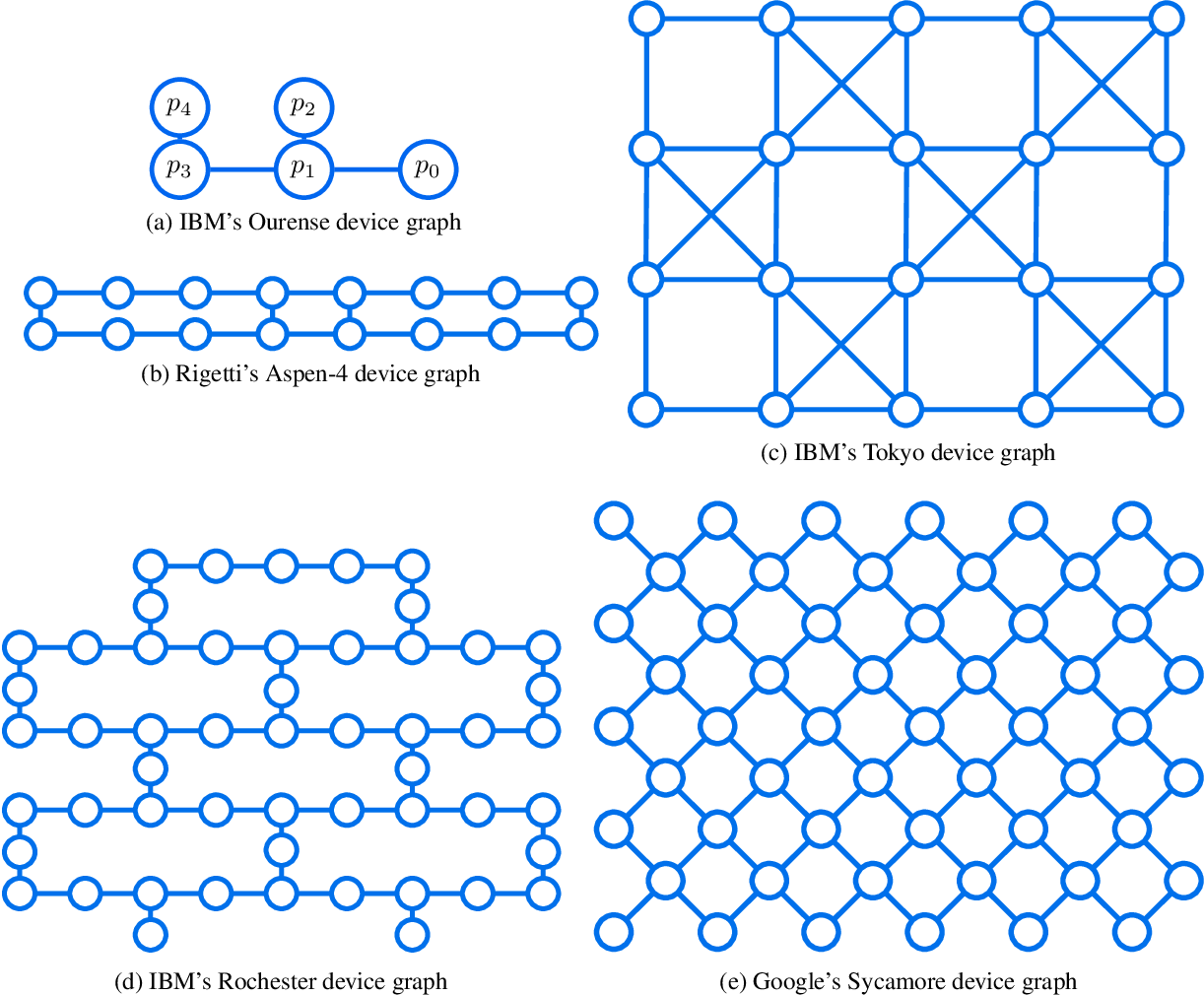}}
\caption{Layout of various superconducting quantum processors (source: \cite{Tan_2021})}
\label{fig:superconducting_layouts}
\end{figure}

Quantum machine learning is an emerging field that has the purpose of exploiting the nature of quantum computing to achieve an advantage over classical machine learning algorithms. In the last years, a lot of research has been done in this sense, of which a theoretical reference can be extensively be found in \cite{Schuld2021}, with contributions also from \cite{Mengoni2019} and \cite{Bartkiewicz2020}.
Given the low number of qubits available as of today in any technology, benchmarks of actual interest cannot be performed and the experiments are limited to problems of very small size, hence making the comparison with the classical counterparts meaningless. Furthermore, the very meaning of quantum advantage itself is questioned, and it is rightfully proposed in \cite{Schuld_2022} that research be not only focused on beating classical algorithms, but also on finding the differences and similarities between the two counterparts and what each one is better at, thus expanding the theory, which is indeed the optimal goal given this early technological stage.\\
Nonetheless, it makes sense to build a foundation that, despite being of little use today, can be the basis of future works. \\
Among the machine learning algorithms that can benefit the most from quantum computing there is the Support Vector Machine algorithm (SVM), of which the most important and expensive step is the computation of a kernel function related to a specific feature map of data into a high-dimensional feature space. When the dimension of the feature space is particularly high, the mapping becomes classically intractable and a quantum computer can be used instead, given the exponential nature of its state space on the number of qubits (the tensor product of $N$ 2-dimensional complex Hilbert spaces $\mathcal{H}=\mathbb{C}^2$ is a $2^N$-dimensional complex Hilbert space $\mathcal{H}_N = (\mathbb{C}^2)^N$). The advantage lies particularly in the fact that the quantum computer can be used directly for obtaining the inner products that make up the kernel, with no need to give the entire high-dimensional state to the training algorithm (performed on a classical computer), but rather the inner products that are priorly calculated on the quantum computer. Naturally, this is only useful when the data themselves actually require being mapped on such a large space to be separated, so this is the scenario considered in this paper. \\
When designing a quantum feature map, it is important to observe that, if the unitary evolution in a quantum circuit could be decomposed as a tensor product of 1-local (single-qubit) unitary operations, then it would be easy to simulate it on a classical computer, since there would be no need to store the entire statevector $\ket{\Psi}$ but rather the state of each single qubit $\ket{q_i}$, since $\ket{\Psi}$ would be simply equal to \begin{equation*}
    \ket{\Psi} = \bigotimes \limits_{i=N}^{1} \ket{q_i}. 
\end{equation*} What would force to store the whole statevector, thus making the classical simulation intractable at a high number of qubits, are 2-local operations, i.e. entangling gates, of which the $CX$ gate is the most frequently used. Indeed, it is easy to show that entangled states cannot be expressed as a tensor product of single-qubit states, taking any Bell state as a simple example, such as $\ket{\Phi^+} = \frac{1}{2} (\ket{0}\otimes\ket{0}+\ket{1}\otimes\ket{1})$, for which $\nexists\; \alpha,\beta,\gamma,\delta\in\mathbb{C} : \ket{\Phi^+} = (\alpha\ket{0}+\beta\ket{1})\otimes(\gamma\ket{0}+\beta\ket{1})$.
In \cite{havlicek_supervised_2019}, a method for generating a dataset separable in a high-dimensional space and a circuit for implementing the feature map is proposed, using the gate model which is native of IBM superconducting quantum computers, considering data with $2$ features, thus needing only $2$ qubits. In this case, the qubit layout of the quantum processor is not a problem, because any layout has at least $2$ qubits that can be directly entangled. However, as the number of features grows and consequently the number of qubits too, the necessity to implement 2-qubit gates on all pairs of qubits becomes a problem, since the layouts of superconducting processors do not allow such a direct connection between qubits because of hardware compatibility issues (see Fig. \ref{fig:superconducting_layouts}). This translates into the necessity to use many swap gates to indirectly make the entanglement between any two qubits possible, which highly increases the depth of the circuit, exposing the circuit more to decoherence.

This problem is much more solvable on  {neutral atom quantum computers}, where the fundamental elements are Rubidium atoms and the position of qubits is not fixed but can be arbitrary, subject to constraints that will be explored in the following sections, and also discussed in the literature (an introduction to neutral atom devices can be found in \cite{henriet_quantum_2020}, whereas a further analysis of the arrangement constraints, applied to Quadratic Unconstrained Binary Optimization problems, can be found in \cite{inproceedings}). It is therefore natural to see neutral atom devices as an ideal alternative to superconducting ones, given the possibility to arrange the qubits with more freedom and, therefore, allowing for more direct connections, reducing the depth of the circuit.

In the literature there is a single example of using this technology for computing quantum kernels (see \cite{albrecht2022quantum}), but it considers graph-structured data and it uses an analogue approach, where the multi-qubit Hamiltonian acting on the entire register is obtained as a result of an optimization process depending on the training set, and the gate model is not used.
In this paper, a different and universal paradigm is presented, where the gate model is used to directly implement the circuits proposed in \cite{havlicek_supervised_2019} on neutral atom devices instead, taking advantage of their arbitrary connectivity. Pasqal quantum computers will be taken as a reference without loss of generality, with a notable alternative being QuEra. An introduction to Pasqal devices can be \cite{Henriet_2020}, whereas \cite{Meet_QuEra_2022} can be consulted as a first read about QuEra. A number equal to $3$ features will be considered, because of the computational effort that would be required for simulating circuits with more qubits since the Pasqal simulator is based on QuTiP (see \cite{JOHANSSON20131234}); however, the process can be generalized to any number of qubits.\\
The rest of the paper is organized as follows.
In subsection II.A, it is first shown how to obtain the main 1-qubit and 1-qubit gates starting from low-level pulses and generalizing the method presented in \cite{silverio_pulser_2022}, that only implements the $R_y$ and $CZ$ gates. The general algorithms proposed in this paper allow to obtain a layer of abstraction where the process of designing a sequence of pulses is replaced by the direct realization of a parameterized gate circuit.
In subsection II.B it is discussed how to exploit the arrangement possibilities given by neutral atom devices in a general framework, even though only $3$ qubits are considered in this work.
In subsections II.C and II.D, then, the formalisms of SVM and Quantum Kernel Estimation (QKE) are briefly introduced.
In subsection II.E the experimental setup is presented in detail.
In section IV the results are shown, and in section V they are discussed to provide the basis for future works.

\section{Methodology}

\subsection{Quantum gate design in neutral atom quantum computers}

\begin{figure}[htbp]
\centerline{\includegraphics[width=0.50\textwidth]{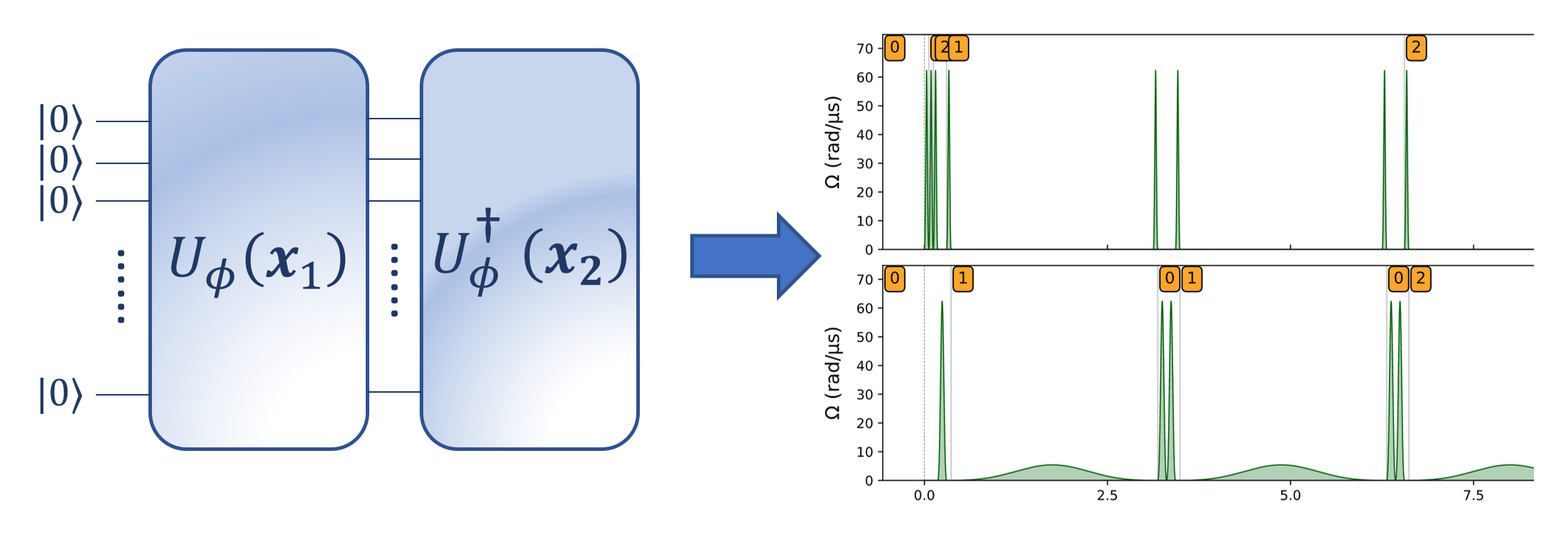}}
\caption{The process of converting a gate circuit into a sequence of pulses for a neutral atom device is achieved using the algorithms presented in this work, used for the specific case of Quantum Kernel Estimation.}
\label{fig:Conversion}
\end{figure}

In this subsection, since there are currently no functions that implement gates on Pasqal quantum computers, a general method for deriving 1-qubit and 2-qubit gates from pulses is shown, and the algorithms corresponding to the primitives for obtaining each gate are illustrated. The algorithms shown in this section will be used in this paper for converting a gate circuit into a sequence of pulses, as in Fig. \ref{fig:Conversion}. When calling a function, if the name of the parameter is not clear from context it is indicated with the notation $parameter \leftarrow parameterValue$.

\subsubsection{Single-qubit gates}
It is first shown how to obtain single-qubit gates and then it is derived how to obtain the $CZ$ and the $CX$ gates, allowing to achieve a universal gate-based toolset.\\
An arbitrary single-qubit unitary rotation can be expressed as a Z-X-Z rotation by means of Euler angles. This makes it possible to represent the rotation as

\begin{equation}\label{eq:u_zxz}
    U\left(\gamma,\theta,\phi\right) = 
    R_z(\gamma)R_x(\theta)R_z(\phi),
\end{equation}
and the reason for using a Z-X-Z rotation will be evident later, following the approach presented in \cite{silverio_pulser_2022}.
\\
Since
\begin{equation}\label{eq:rz_def}
    R_z(\phi) = \begin{bmatrix}
        e^{-i\frac{\phi}{2}} & 0\\
        0 & e^{i\frac{\phi}{2}}
    \end{bmatrix}
\end{equation}
and 
\begin{equation}\label{eq:rx_def}
    R_x(\theta) = \begin{bmatrix}
        \cos{\frac{\theta}{2}} & -i\sin{\frac{\theta}{2}} \\
        i\sin{\frac{\theta}{2}} & \cos{\frac{\theta}{2}}  
    \end{bmatrix},
\end{equation}
then $U$ is equal to
\begin{equation}\label{eq:u_fullmatrix}
    U\left(\gamma,\theta,\phi\right) = \begin{bmatrix}
        e^{-i\frac{\phi+\gamma}{2}}\cos{\frac{\theta}{2}} & -ie^{i\frac{\phi-\gamma}{2}}\sin{\frac{\theta}{2}} \\
        -ie^{i\frac{\gamma-\phi}{2}}\sin{\frac{\theta}{2}} & e^{i\frac{\phi+\gamma}{2}}\cos{\frac{\theta}{2}}
    \end{bmatrix}.
\end{equation}
\\
Up to global phase, i.e. up to a multiplication by a scalar $e^{i\lambda}, \lambda\in\mathbb{R}$ that doesn't make any actual difference on the representation of a quantum state, these equalities hold:
\begin{equation}\label{eq:rz_from_u}
    R_z(\phi) = U\left(\frac{\phi}{2},0,\frac{\phi}{2}\right),
\end{equation} 
\begin{equation}\label{eq:rx_from_u}
    R_x(\theta) = U\left(0,\theta,0\right).
\end{equation}
\\
The Hadamard gate, represented by its matrix
\begin{equation*}\label{eq:h_def}
    H = \frac{1}{\sqrt{2}}\begin{bmatrix}
        1 & 1 \\
        1 & -1
    \end{bmatrix},
\end{equation*}
can be obtained from $U$ as 
\begin{equation*}\label{eq:h_from_u}
    H = U\left(\frac{\pi}{2},\frac{\pi}{2},\frac{\pi}{2}\right).
\end{equation*}
\\
Similarly, the $Z$ gate and the $X$ gate, respectively represented by their Pauli matrices
\begin{equation*}\label{eq:z_def}
    \sigma_Z = \begin{bmatrix}
        1 & 0 \\
        0 & -1
    \end{bmatrix},
\end{equation*}
\begin{equation*}\label{eq:x_def}
    \sigma_X = \begin{bmatrix}
        0 & 1 \\
        1 & 0
    \end{bmatrix},
\end{equation*}

can both be obtained from $U$, using Eq. \ref{eq:rz_from_u} and \ref{eq:rx_from_u}, as
\begin{equation*}\label{eq:z_from_u}
    \sigma_Z = R_z(\pi) = U\left(\frac{\pi}{2},0,\frac{\pi}{2}\right),
\end{equation*}
\begin{equation*}\label{eq:x_from_u}
    \sigma_X = R_x(\pi) =  U\left(0,\pi,0\right). 
\end{equation*}
\\
Focusing on Pasqal neutral atom quantum computers, where Rubidium atoms are used, in \cite{silverio_pulser_2022} it is shown that there are generally two different channels, the  {Rydberg channel} and the  {Raman channel}. The Raman channel, also referred to as the digital channel, allows for the discrete manipulation of single qubits, allowing to realize single-qubit gates. There are two possible states, the  {ground state} $\ket{g}$ and the  {hyperfine state} $\ket{h}$, respectively corresponding to $\ket{0}$ and $\ket{1}$. A pulse can drive the transition between these two levels of a qubit by means of the drive Hamiltonian \begin{equation}
    H^D (t) = \frac{\hbar}{2} (\Omega(t)\cos(\phi),-\Omega(t)\sin(\phi),-\delta(t)) \cdot (\sigma_X, \sigma_Y, \sigma_Z)
\end{equation}
or, more compactly,
\begin{equation}
    H^D (t) = \frac{\hbar}{2} \boldsymbol{\Omega}(t) \cdot \boldsymbol{\sigma},
\end{equation}
where $\Omega(t)$ is the Rabi frequency, i.e. the amplitude of the signal measured in $rad\;\mu s^{-1}$, $\phi$ is the phase of the pulse 
and $\delta(t) = \omega(t) - \frac{|E_a - E_b|}{\hbar}$, being $\omega(t)$ is the frequency of the signal and $E_a,\;E_b$ the two energy levels of the qubit.
In this paper only resonant pulses ($\delta = 0$) of duration $T$ and phase $\phi$ will be considered, so that $\boldsymbol{\Omega}(t) = (\Omega(t)\cos(\phi),-\Omega(t)\sin(\phi),0)$, generating a rotation along the axis $\boldsymbol{\hat{u}}(\phi) = \left(\cos{\phi},-\sin{\phi},0\right)$ of the Bloch sphere, by an angle equal to
\begin{equation}\label{eq:theta_from_omega}
    \theta = \int_0^T{\Omega(t)dt},
\end{equation}
which corresponds to the following unitary
\begin{equation}
    R_{\hat{u}_{\phi}}(\theta)=\cos\left(\frac{\theta}{2}\right)\mathbb{I}-i\sin\left(\frac{\theta}{2}\right)(\boldsymbol{\hat{u}} \cdot \boldsymbol{\sigma}),
\end{equation}
and this corresponds to
\begin{equation}\label{eq:rotation_equator}
    R_{\hat{u}_{\phi}}(\theta)=R_z(-\phi)R_x(\theta)R_z(\phi).
\end{equation}
\\
The difference between Eq. \ref{eq:rotation_equator} and Eq. \ref{eq:u_zxz} lies in the fact that the leftmost operation in the former is $R_z(-\phi)$, whereas in the latter it is $R_z(\gamma)$.
This is irrelevant if this rotation is performed right before measurement, whereas, if other rotations are performed later, a {phase shift} equal to $\gamma + \phi$ has to be added to the channel, so that the next rotation $R_{\hat{u}_{\phi_2}}(\theta_2)$ will consider the phase shift, i.e.
\begin{equation*}
    R_{\hat{u}_{\phi_2}}(\theta_2) = R_z(-\phi_2-\gamma-\phi)R_x(\theta_2)R_z(\phi_2+\gamma+\phi),
\end{equation*}
which, after the first rotation, corresponds to
\begin{equation*}
    [R_z(-\phi_2-\gamma-\phi)R_x(\theta_2)R_z(\phi_2)][R_z(\gamma)R_x(\theta)R_z(\phi)],
\end{equation*}
and this process can be reiterated.

\begin{figure}[htbp]
\centerline{\includegraphics[width=0.5\textwidth]{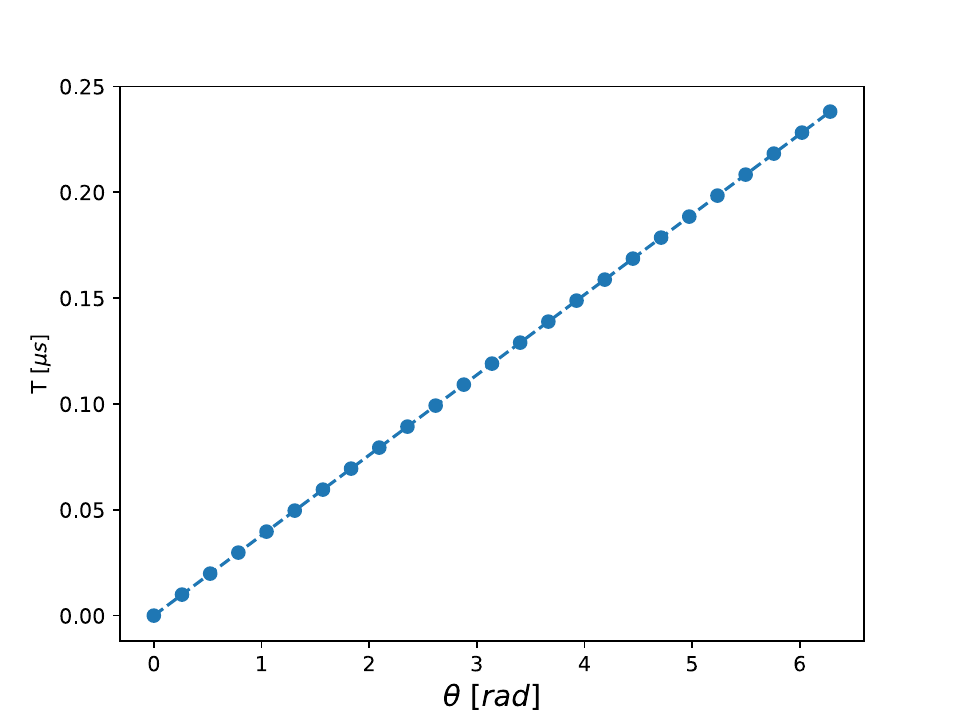}}
\caption{A plot showing the duration $T$ of a pulse corresponding to a single-qubit rotation $U(\gamma,\theta,\phi)$ depending on $\theta$ on the Chadoq2 device ($A=62.83\;rad\;\mu s^{-1}$).}
\label{fig:durations}
\end{figure}

As for the amplitude of the pulse $\Omega(t)$, the Blackman window function is chosen so that the spectral leakage is minimized. This function, with maximum amplitude $A$ [$rad/ \mu s$] and in discrete time, is defined as
\begin{equation}
    \Omega(t) = A\left(0.42 - 0.5\cos\left(\frac{2\pi t}{T}\right) + 0.08\cos\left(\frac{4\pi t}{T}\right)\right),
\end{equation}
where $t = n\Delta t$, $T = N\Delta t$ and $\Delta t$ is the time step. As $\Delta t \rightarrow 0$, the area under $\Omega(t)$, which in turn is the rotation angle $\theta$, can be calculated as
\begin{equation*}
    \theta = \int_0^T \Omega(t) dt = 0.42 A T \space,
\end{equation*}
which means that, to get an angle $\theta$ out of a pulse with maximum amplitude $A$, $T$ must be equal to 
\begin{equation}\label{eq:T}
    T = \frac{\theta}{0.42A}.
\end{equation}
The maximum amplitude $A$ is the maximum output of the driving channel and it depends both on the chosen device and channel; in this paper, the Chadoq2 device was used, with $A = 62.83\;rad\;\mu s^{-1}$. A plot of $T$ depending on $\theta$ is shown in Fig. \ref{fig:durations}, with this value of $A$.

\begin{algorithm}
\caption{$U_{ZXZ}$}\label{alg:cap}
\SetKwInOut{Parameters}{Parameters}
\Parameters{$\gamma$, $\theta$, $\phi$, max, sequence, channel, qubit}

    channel.target $\leftarrow$ qubit\\
    $\gamma \leftarrow \gamma \mod 2\pi$\\
    $\theta \leftarrow \theta \mod 2\pi$\\
    $\phi \leftarrow \phi \mod 2\pi$\\

    \uIf{$\theta \neq 0$}{
        blackman $\leftarrow$ Blackman(\\
        \qquad max\_amplitude $\leftarrow$ max, \\
        \qquad area $\leftarrow$ $\theta$)\\
        pulse $\leftarrow$ Pulse(\\
        \qquad waveform $\leftarrow$ blackman, \\
        \qquad detuning $\leftarrow$ 0, \\
        \qquad phase $\leftarrow$ $\phi$, \\
        \qquad post\_phase\_shift $\leftarrow$ $(\gamma+\phi)\mod(2\pi))$\\
        sequence.add(pulse, channel) \\
    }
    
    \Else{
        sequence.phase\_shift(shift $\leftarrow$ $\gamma + \phi$, channel, qubit)\\
    }
    
    \KwRet{sequence} 
\end{algorithm}

The other gates can be obtained subsequently, as shown in the various algorithms in this paper.
\begin{algorithm}
\caption{$R_X$}\label{alg:rx}
\SetKwInOut{Parameters}{Parameters}
\Parameters{$\theta$, max, sequence, channel, qubit}
    \KwRet{$U_{ZXZ}$(0, $\theta$, 0, max, sequence, channel, qubit)} 
\end{algorithm}

\begin{algorithm}
\caption{$R_Z$}\label{alg:rz}
\SetKwInOut{Parameters}{Parameters}
\Parameters{$\phi$, sequence, channel, qubit}
    sequence.phase\_shift(shift $\leftarrow$ $\phi$, channel, qubit)\\
    \KwRet{sequence} 
\end{algorithm}
\begin{algorithm}
\caption{$X$}\label{alg:x}
\SetKwInOut{Parameters}{Parameters}
\Parameters{max, sequence, channel, qubit}
    \KwRet{$R_X$($\pi$, max, sequence, channel, qubit)} 
\end{algorithm}
\begin{algorithm}
\caption{$H$}\label{alg:h}
\SetKwInOut{Parameters}{Parameters}
\Parameters{sequence, channel, qubit}
    \KwRet{$U_{ZXZ}$($\frac{\pi}{2}$,$\frac{\pi}{2}$,$\frac{\pi}{2}$, max, sequence, channel, qubit)} 
\end{algorithm}

\subsubsection{Two-qubit gates}
In neutral atom devices, what allows entanglement and, in turn, two-qubit gates, is the  {Rydberg blockade}, which is not possible considering only the $\ket{g}$ and $\ket{h}$ states; hence, the Raman channel cannot be used for this purpose. 
Using the Rydberg channel, instead, a transition between the ground state $\ket{g}$ and the  {Rydberg state} $\ket{r}$ can be driven, in addition to the hyperfine state $\ket{h}$, obtaining a 3-level system. In this configuration, the global Hamiltonian becomes
\begin{equation}
    \mathcal{H}(t) = \sum \limits_{i} \left(H_i^D (t)+\sum \limits_{j<i}\frac{C_6}{R_{ij}^6}\hat{n}_i\hat{n}_j\right),
\end{equation}
where to the drive Hamiltonian of each qubit is added a term that accounts for the van der Waals forces between each pair of atoms. For the device Chadoq2, $C_6\hbar^{-1} = 5008\;GHz\;\mu m^6$. 

\begin{figure}[htbp]
\centerline{\includegraphics[width=0.5\textwidth]{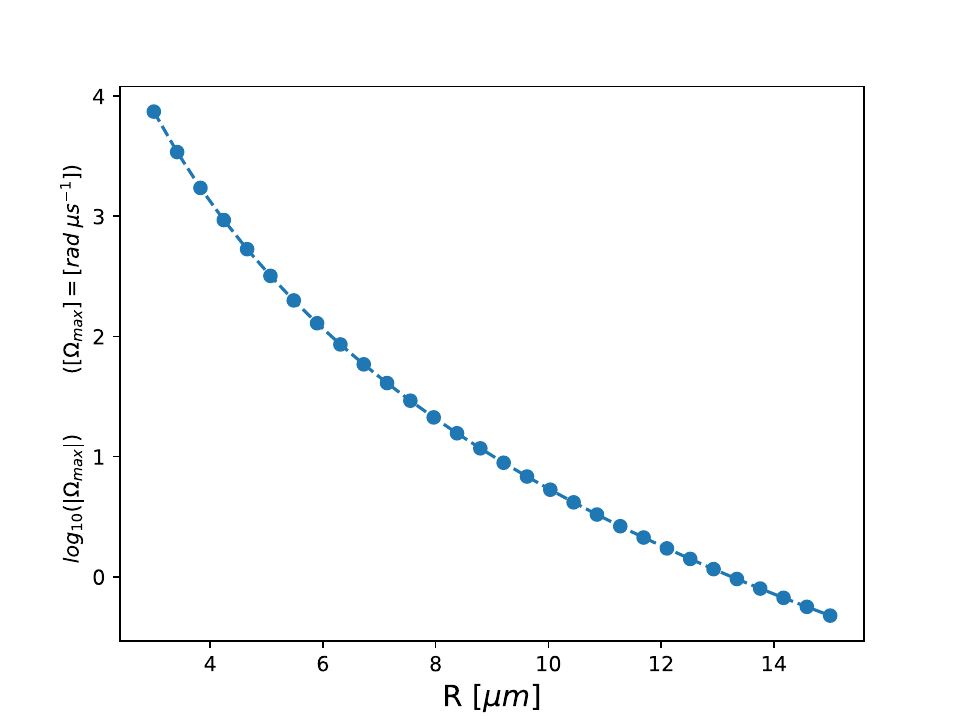}}
\caption{A plot showing the $log_{10}$ of the value of $\Omega_{max}$ with respect to the atom distance $R$ for making the Rydberg blockade possible.}
\label{fig:rabimax}
\end{figure}

Furthermore, $R_{ij}$ is the distance between the $i$-th and the $j$-th atoms, and $\hat{n}_i$ is the projector $\ket{r}\bra{r}_i$ acting on the $i$-th atom. When a pulse with $\delta=0$ and maximum amplitude $\Omega_{max}$ is driven through the Rydberg channel, the entanglement between two atoms at distance $R$ can happen only if 
\begin{equation}
    R \ll R_b = \left(\frac{C_6}{\hbar \Omega_{max}}\right)^{\frac{1}{6}},
\end{equation} being $R_b$ the Rydberg blockade radius, which corresponds to a constraint on $\Omega_{max}$ given $R$. The $\pi-2\pi-\pi$ sequence proposed in \cite{silverio_pulser_2022} can be used for obtaining a $CZ$ gate with global phase $-1$, where the considered states are still $\ket{g} = \ket{0}$ and $\ket{h} = \ket{1}$, while $\ket{r}$ is used only for implementing the $CZ$. However, since the $CX$ gate is needed too, the sequence must be modified. By observing that $X = HZH$, i.e. $Z\sim X$ with respect to the Hadamard basis $\{\ket{+},\ket{-}\}$, it is sufficient to add a pulse corresponding to the $H$ gate on the target qubit right before and after the $2\pi$ pulse, using the Raman channel. On the Rydberg channel, the constraint on the maximum $\Omega$ depending on $R$ is only applied for the $2\pi$ pulse. In Fig. \ref{fig:rabimax}, a plot of the logarithm base 10 of $\Omega_{max}$ depending on $R$ for obtaining the Rydberg blockade is shown.
A detailed explanation of the Blockade effect can be found in the literature (see \cite{Zhang2010}, \cite{Wilk2010}, \cite{Graham2019}).

\begin{algorithm}
\caption{$CX$}\label{alg:rz}
\SetKwInOut{Parameters}{Parameters}
\Parameters{sequence, rydbergChannel, ramanChannel, max, maxRyd, controlQubit, targetQubit}
    $\pi$\_wave $\leftarrow$ Blackman(\\
    \qquad max\_amplitude $\leftarrow$ max, \\
    \qquad area $\leftarrow$ $\pi$)\\
    $2\pi$\_wave $\leftarrow$ Blackman(\\
    \qquad max\_amplitude $\leftarrow$ maxRyd, \\
    \qquad area $\leftarrow$ $2\pi$)\\
    $\pi$\_pulse $\leftarrow$ Pulse(\\
    \qquad waveform $\leftarrow$ $\pi$\_wave, \\
    \qquad detuning $\leftarrow$ 0, \\
    \qquad phase $\leftarrow$ 0)\\
    $2\pi$\_pulse $\leftarrow$ Pulse(\\
    \qquad waveform $\leftarrow$ $2\pi$\_wave, \\
    \qquad detuning $\leftarrow$ 0, \\
    \qquad phase $\leftarrow$ 0)\\
    rydbergChannel.target $\leftarrow$ controlQubit \\
    sequence.add($\pi$\_pulse,rydbergChannel) \\
    ramanChannel.target $\leftarrow$ targetQubit \\
    sequence $\leftarrow$ H(sequence,rydbergChannel,targetQubit) \\
    rydbergChannel.target $\leftarrow$ controlQubit \\
    sequence.add($2\pi$\_pulse,rydbergChannel) \\
    sequence $\leftarrow$ H(sequence,rydbergChannel,targetQubit)\\
    rydbergChannel.target $\leftarrow$ controlQubit \\
    sequence.add($\pi$\_pulse,rydbergChannel) \\
    \KwRet{sequence} 
    
\end{algorithm}

\subsection{Arrangement of neutral atoms}
\begin{figure}[htbp]
\centerline{\includegraphics[width=0.4\textwidth]{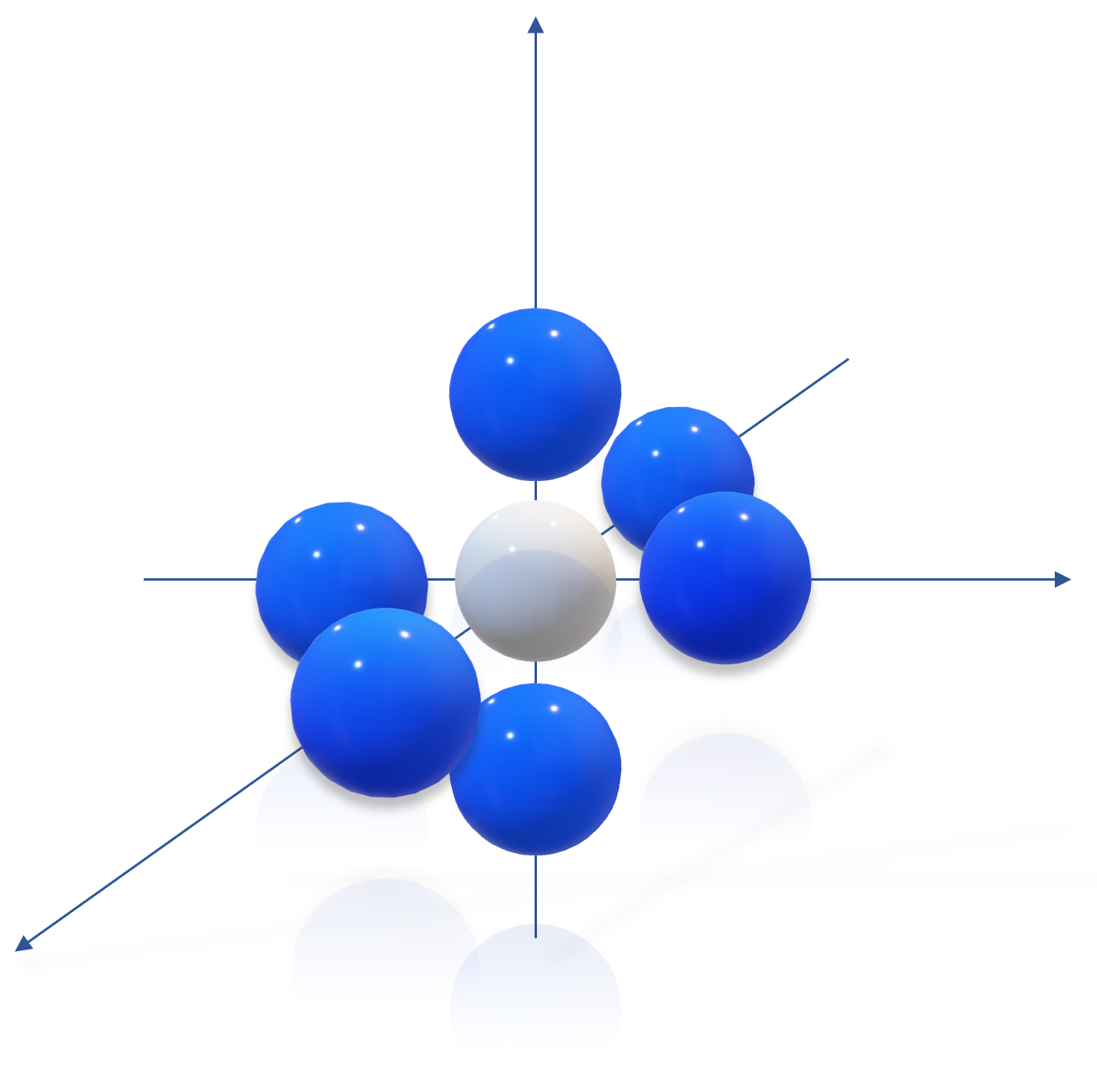}}
\caption{A possible pattern of arranging the neighbors of each atom, if a fixed distance $d$ is wanted. In 3D space, only 6 neighbors are possible with this constraint.}
\label{fig:3d_atoms}
\end{figure}

If the chosen device allows to position the atoms in three-dimensional space and not just on a plane, a lattice could be ideally created where the innermost atoms are arranged as in Fig. \ref{fig:3d_atoms}.

\begin{figure}[htbp]
\centerline{\includegraphics[width=0.5\textwidth]{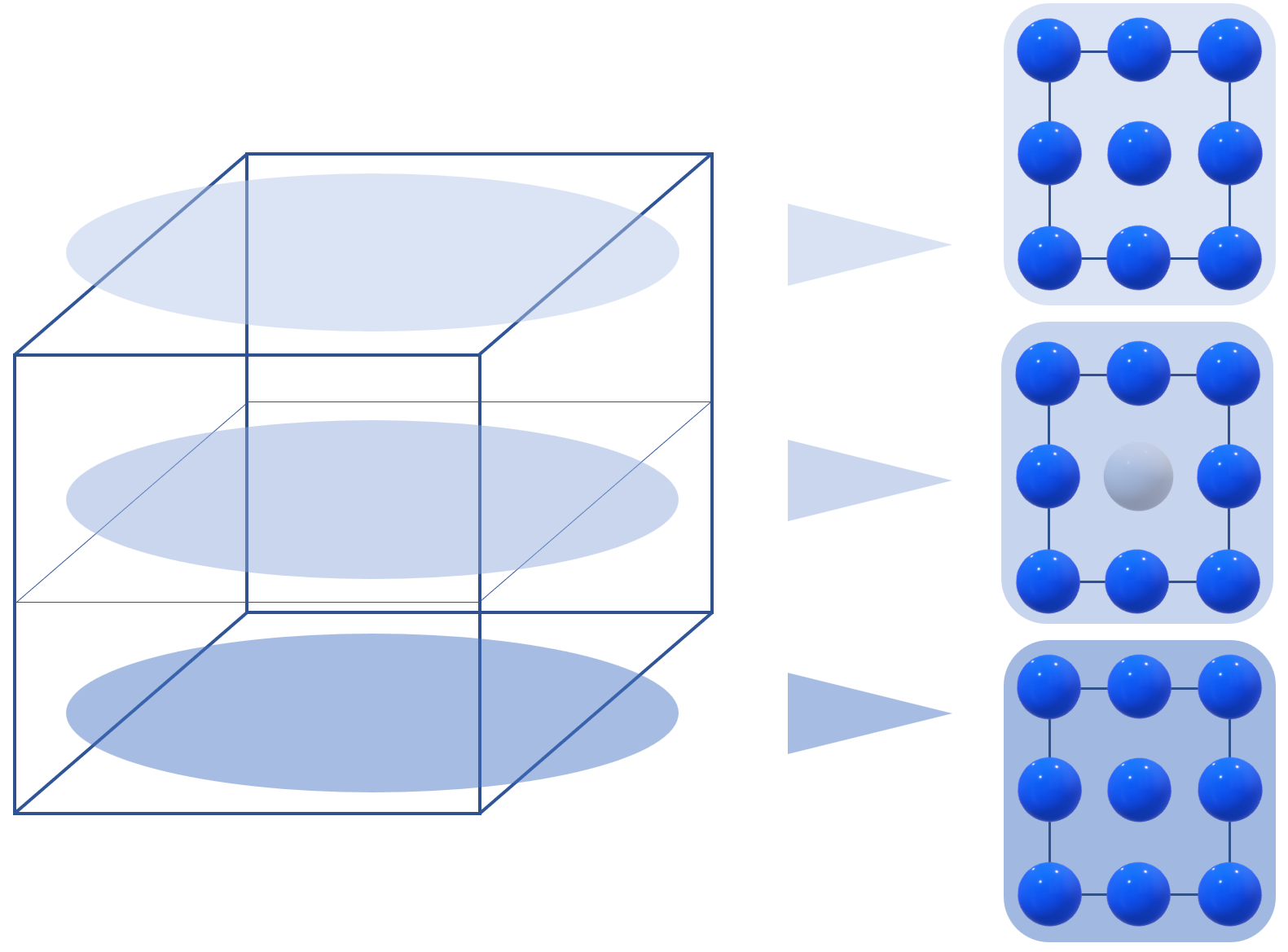}}
\caption{Another possible pattern of arranging the neighbors of each atom. A cube of length $2d$ is considered, and the top face, the bottom face and the face resulting from a central cut are expanded on the right to show the atom arrangements. The central atom (in white) is at a minimum distance $d$ and at a maximum distance $d\sqrt{3}$ from all its neighbors. With this pattern, it is possible to have a total of $26$ neighbors for each atom. }
\label{fig:3d_cube}
\end{figure}

 In particular, assuming that a fixed distance $d$ between atoms is wanted, it is clear that, considering a cube of length $2d$ and an atom placed at its center, all of the atoms that are at the center of each of the $6$ faces of the cube will be at distance $d$ from it. This pattern can be repeated in space, hence creating a lattice of atoms where, inside of it, each atom is at distance $d$ from its $6$ neighbors. If it is not important that the distance be constant, it can be seen that, building a lattice following the pattern in Fig. \ref{fig:3d_cube}, the central atom (white) will have a minimum distance of $d$ and a maximum distance of $d\sqrt{3}$ from its neighbors, with the advantage that now it has $26$ neighbors instead of just $6$, which allows for direct entanglement between the central qubit and all the other qubits in that neighborhood without using swap gates, which is a noticeable reduction considering the fact that, with superconducting devices, a qubit can generally be put in direct entanglement with at most $2$ other qubits. Naturally, the minimum distance requirement of the device and the maximum distance according to the Rydberg blockade radius must be respected. In this paper, however, a device that only allows planar arrangement is considered, because as of today there is no possibility to simulate a device with spatial arrangement, but the same arguments hold for either scenarios.

\subsection{Support Vector Machines}
In supervised binary classification, the objective is to determine whether a given sample belongs to one class or the other. SVM is a machine learning technique that consists in finding a separating hyperplane between the samples of the two distributions.\\
Formally, each sample with $f$ features can be seen as a vector $\boldsymbol{x} \in \mathcal{X} \subset \mathbb{R}^f$, paired with a label $y \in \mathcal{Y} = \{+1,-1\}$. The training set is therefore the set of all points \footnote{$[n]=\{1,\dots,n\}$}
\begin{equation*}
    M = \{\boldsymbol{\omega_{M,i}}=(\boldsymbol{x}_{M,i},y_{M,i}) \in \Omega = \mathcal{X} \times \mathcal{Y}\}_{i\in[m]},
\end{equation*}
used in the training phase, whereas the test set is
\begin{equation*}
    S = \{\omega_{S,i} \in \Omega\}_{i\in[s]},
\end{equation*}
used in the testing phase.

The training step consists in solving the following optimization problem,
\begin{equation}
\begin{gathered}
    \operatorname*{min}_{\boldsymbol{w},b} \left(\lambda||\boldsymbol{w}||^2+L_S^{hinge}((\boldsymbol{w},b))\right) ,
\end{gathered}
\end{equation}
where $L_S^{hinge}((\boldsymbol{w},b))$ is the average of the hinge loss $l^{hinge}((\boldsymbol{w},b),(\boldsymbol{x},y))$ over all the samples $(\boldsymbol{x}_M,y_M)$ of the training set, and the hinge loss is defined as
\begin{equation}
\begin{gathered}
    {l}^{hinge}((\boldsymbol{w},b),(\boldsymbol{x},y)) = \operatorname*{max} {(0,1-y(\langle\boldsymbol{w},\boldsymbol{x}\rangle+b))}
\end{gathered}
\end{equation}

so that the resulting parameters define the linearly separating hyperplane (see \cite{shalev2014understanding} for an in-depth derivation). After this step, a classifier $f$ can be derived such that $f: \mathbb{R}^f \rightarrow \mathcal{Y}\cup\{0\}$, where the value $0$ can be associated with the label $+1$. \\
However, since data is frequently non-linearly separable, a nonlinear function is applied to data in order to account for nonlinearities, which results in an increased number of dimensions. This is allowed by means of a mapping $\Phi : \mathcal{X} \rightarrow \mathcal{F}$, where $\mathcal{F}$ is a Hilbert space called the \textit{feature space}. 
Since the mapping can be computationally hard to calculate for each sample if $\mathcal{F}$ has a high dimension, the \textit{kernel trick} is used. Defining the \textit{kernel} as a function $K: \mathcal{X}\times\mathcal{X} \rightarrow \mathbb{R}$, where \begin{equation}
    K(\boldsymbol{x}_a,\boldsymbol{x}_b) = \langle\Phi(\boldsymbol{x}_a),\Phi(\boldsymbol{x}_b)\rangle,
\end{equation}
it is often the case that, given two points, their kernel, i.e. the scalar product of their feature maps, is much easier to calculate than their respective feature maps separately.
Due to the \textit{representer theorem} (see \cite{shalev2014understanding}), there exists an $\boldsymbol{\alpha}$ such that the optimal solution of the problem can be written as
\begin{equation}
    \boldsymbol{w}^* = \sum\limits_{i=1}^M \alpha_i \Phi(\boldsymbol{x}_{M,i})
\end{equation} and, introducing $K_M \in \mathbb{R}^{m,m}$ as the \textit{training kernel matrix} defined as
\begin{equation}
    (K_M)_{i,j} = K(\boldsymbol{x}_{M,i},\boldsymbol{x}_{M,j}),
\end{equation} it can be derived that

\begin{equation}
    \langle\boldsymbol{w},\psi(\boldsymbol{x}_{M,i})\rangle = (K_M \boldsymbol{\alpha})_i
\end{equation} and
 
\begin{equation}
\|\boldsymbol{w}\|^2 = \boldsymbol{\alpha}^T K_M \boldsymbol{\alpha},    
\end{equation}

so that the optimization problem is now expressed as
\begin{equation}\label{eq:optimization}
\begin{gathered}
    \operatorname*{min}_{\boldsymbol{\alpha},b} \left({\lambda\boldsymbol{\alpha}^T K_M \boldsymbol{\alpha}+\frac{1}{m}\sum\limits_{i=1}^{m} \max{(0,1-y_i (K_M \boldsymbol{\alpha})_i}+b)}\right).
\end{gathered}
\end{equation}

At this point, to classify an instance $\boldsymbol{x}_{S}$ belonging to the test set, it is sufficient to calculate the quantity
\begin{equation}
    \hat{y}_S = sign\left(\sum\limits_{i=1}^{m} y_{M,i} \alpha_i K(\boldsymbol{x}_{S},\boldsymbol{x}_{M,i})+b\right).
\end{equation} It is therefore natural to calculate the \textit{test kernel matrix} $K_S \in \mathbb{R}^{s,m}$, defined as
\begin{equation}\label{eq:classification}
    (K_S)_{i,j} = K(\boldsymbol{x}_{S,i},\boldsymbol{x}_{M,j}),
\end{equation}
to speed up the evaluation of the accuracy on the test set.

\subsection{Quantum Kernel Estimation}
When the kernel is hard to compute classically, a quantum computer can be used instead, leaving the training part to the classical computer once the kernel is calculated. In particular, a unitary $U_\Phi(\boldsymbol{x})$ can be realized such that, starting from the state $\ket{0} = \ket{0\dots0}$, it performs the feature mapping of $\boldsymbol{x}$ directly on the quantum computer, obtaining $U_\Phi(\boldsymbol{x})\ket{0}$. Following the method presented in \cite{havlicek_supervised_2019}, given a state $\psi$, the density matrices $\ket{\psi}\bra{\psi}$ are considered instead of their statevector $\ket{\psi}$, so that inner products of two different states do not depend on the global phase of their representation. This is why the kernel is calculated as $K(\boldsymbol{x}_1,\boldsymbol{x}_2) = |\braket{\Phi(\boldsymbol{x}_2)|\Phi(\boldsymbol{x}_1)}|^2$. This is equivalent to
\begin{equation}
    K(\boldsymbol{x}_1,\boldsymbol{x}_2) = |\bra{0}U^\dag_\Phi(\boldsymbol{x}_2) U_\Phi(\boldsymbol{x}_1)\ket{0}|^2,
\end{equation}
which is the square of the probability of measuring all $0$ in the Pauli-Z base after performing $U^\dag_\Phi(\boldsymbol{x}_2) U_\Phi(\boldsymbol{x}_1)\ket{0}$.
In other words, the kernel matrix $(K_M)_{i,j} = K(\boldsymbol{x}_{M,i},\boldsymbol{x}_{M,j})$ can be estimated by performing many times each $U^\dag_\Phi(\boldsymbol{x}_{M,j}) U_\Phi(\boldsymbol{x}_{M,i})\ket{0}$ and calculating the square of the counts of $0$ measurements. In the same way, $K_S$ can be estimated. Naturally, the higher the number of runs for each $(i,j)$ couple, the higher the accuracy of the estimation. \\
In summary, the optimization problem is finally expressed by substituing in Eq. \ref{eq:optimization} the $K_M$ that is estimated.

As shown in the previously cited paper, considering the training step and being $|M|$ the cardinality of the training set, a total number of executions equal to $\mathcal{O}(\delta^{-2}|M|^4)$ allows to obtain a training kernel matrix $\hat{K}_M$ that differs in operator norm from the true $K_M$ by at most $\|K_M-\hat{K}_M\|^2\leq\delta$.\\   
Finally, the training and the testing steps of the SVM are performed on a classical computer by using the kernel matrices calculated in the previous step. 
\subsection{Experimental setup}

\begin{figure}[htbp]
\centerline{\includegraphics[width=0.3\textwidth]{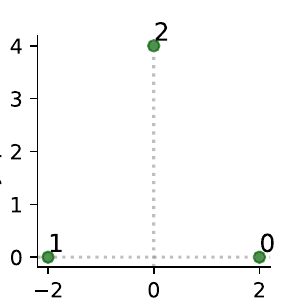}}
\caption{Arrangement of the atoms in this experiment with the Chadoq2 device. The distance $|q_0-q_1|$ is $4\;\mu m$, while the distance $|q_0-q_2|$ and $|q_1-q_2|$ is $4.47\;\mu m$}
\label{fig:register}
\end{figure}

The experiment is performed simulating the Pasqal Chadoq2 device. 
According to \cite{pulserCoreFeatures}, only planar arrangement of atoms is possible, with a maximum of $100$ atoms, a minimum distance of $4 \mu m$ between each pair and a maximum distance from the origin equal to $50 \mu m$. Both the local Rydberg channel and the local Raman channel share the specifications listed in Table \ref{tab1}.

\begin{table}[htbp]
\caption{Chadoq2 local channels specifications}
\begin{center}
\begin{tabular}{|c|c|}
\hline

\textbf{\textbf{\textit{Quantity}}}& \textbf{\textit{Value}}\\
\hline
Maximum $\Omega$ & $62.83\;rad\; \mu s^{-1}$  \\
\hline
Maximum $|\delta|$ & $125.7\;rad\;\mu s^{-1}$\\
\hline
Minimum time between retargets & $220\;ns$\\
\hline

\end{tabular}
\label{tab1}
\end{center}
\end{table}

Since only resonant pulses will be used, i.e. with $\delta = 0$, the constraint about the maximum $|\delta|$ does not interfere with the experiment. It is however worth to notice that every time a channel changes the target qubit there is a loss of at least $0.22 \; \mu s$.
A register of $N=3$ qubits is prepared as in Fig. \ref{fig:register}, with a minimum distance of $4 \mu m$ and a maximum distance of $4.47\;\mu m$ between atoms.

\begin{figure}[htbp]
\centerline{\includegraphics[width=0.5\textwidth]{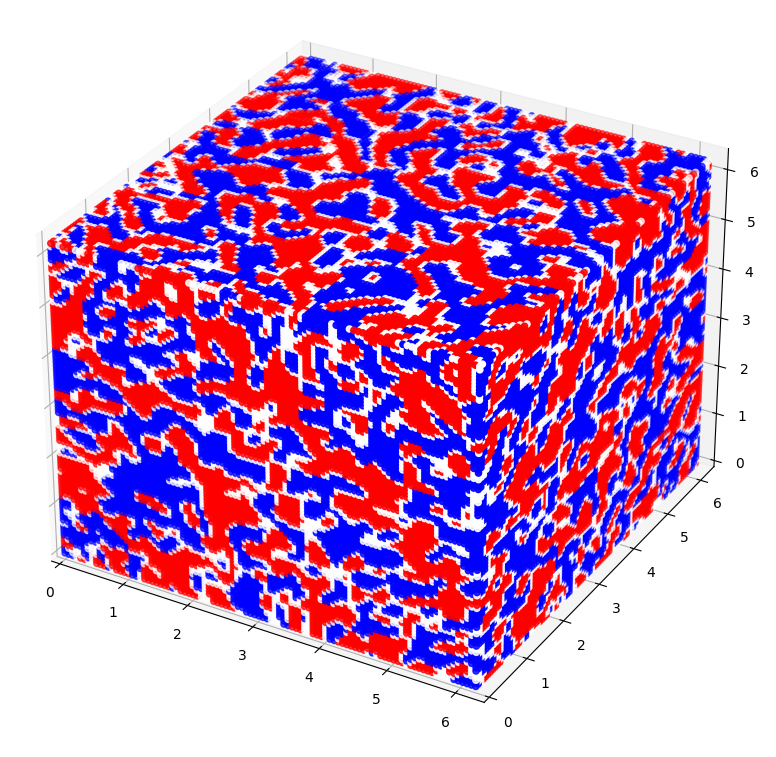}}
\caption{The ground truth regions are shown for the frontier of the $(0,2\pi]^3$ volume. The blue regions are the points $\boldsymbol{x}$ such that $\braket{\Phi(\boldsymbol{x})|V^\dag \boldsymbol{f} V | \Phi(\boldsymbol{x})} \geq \Delta$, whereas the red regions are such that $\braket{\Phi(\boldsymbol{x})|V^\dag \boldsymbol{f} V | \Phi(\boldsymbol{x})} \leq -\Delta$. The separation gap is in white. The function \texttt{ad\_hoc\_data} was modified to return a $100\times100\times100$ uniform grid instead of a $20\times20\times20$ one when $n=3$, to obtain a clearer visualization. Inside the volume, the regions of each section change with continuity.}
\label{fig:regions}
\end{figure}

\begin{figure*}[htbp]
\centerline{\includegraphics[width=\textwidth]{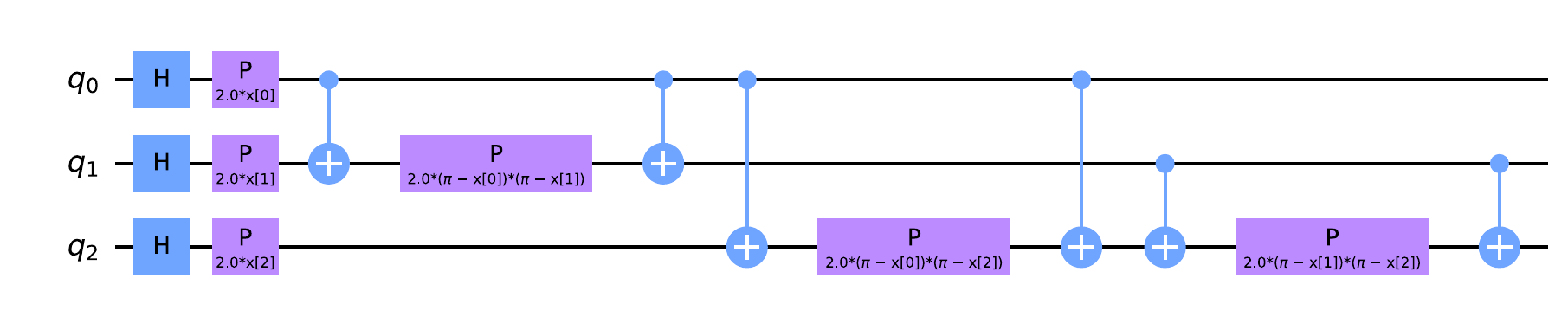}}
\caption{The gate circuit that implements the feature map $U_\phi(\boldsymbol{x})$ with $1$ repetition. The actual circuit uses a $U_\phi(\boldsymbol{x})$ with $2$ repetitions, followed by a $U^\dag_\phi(\boldsymbol{x})$.}
\label{fig:qiskit_featuremap}
\end{figure*}

\begin{figure*}[htbp]
\centerline{\includegraphics[width=1.1\textwidth]{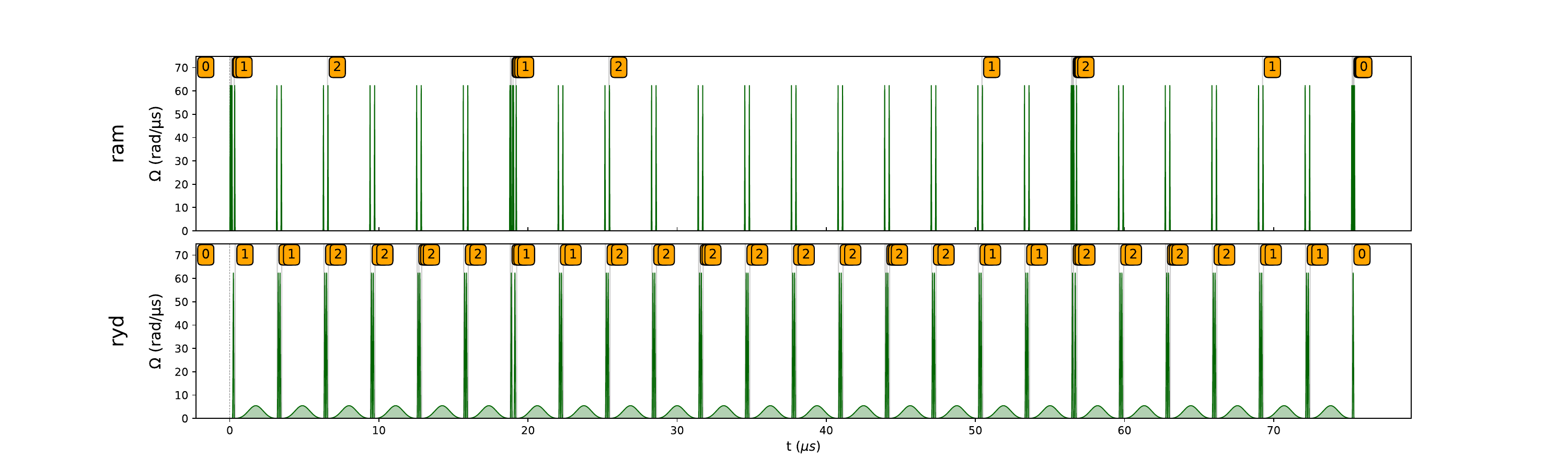}}
\caption{The resulting sequence of pulses corresponding to the QKE circuit ($U_\phi(\boldsymbol{x}U^\dag_\phi(\boldsymbol{x})$). The channel at the top is the local Raman channel, whereas the channel at the bottom is the local Rydberg channel. The highlighted numbers indicate the target qubit of each pulse, that changes at each transition.}
\label{fig:seq}
\end{figure*}

The dataset is generated using Qiskit's function \verb|ad_hoc_data| inside the package \verb|qiskit_machine_learning.datasets|, with a seed equal to $10000$, and with a training set size equal to $20$, a test set size equal to $10$, $3$ features and a gap equal to $\Delta = 0.1$. Since there are two possible classes, the function actually generates a dataset with $m = |M| = 40$ training samples and $s = |S| = 20$ test samples, where half samples are of one class and half samples are of the other one.

More formally, the dataset is prepared by choosing a unitary $V \in SU(2^3)$ and $\boldsymbol{f}=Z_1 Z_2 Z_3$, so that the ground truth labels of each sample are assigned in such a way that, given a randomly generated sample $\boldsymbol{x} \in (0,2\pi]^3$, its label is $1$ if $\braket{\Phi(\boldsymbol{x})|V^\dag \boldsymbol{f} V | \Phi(\boldsymbol{x})} \geq \Delta$, while if $\braket{\Phi(\boldsymbol{x})|V^\dag \boldsymbol{f} V | \Phi(\boldsymbol{x})} \leq -\Delta$ the label $-1$ is assigned (the unitary $\Phi(\boldsymbol{x})$ is the same used in the QKE circuit for performing the feature mapping). To give and idea of the complexity of the separation of the classes, in Fig. \ref{fig:regions} the ground truth regions are shown for the frontier of the $(0,2\pi]^3$ volume, where red corresponds to label $-1$, blue to $+1$ and white to the separation gap.

To perform the feature mapping on the neutral atom device, the circuit generated by the instruction \verb|ZZFeatureMap| inside the package \verb|qiskit.circuit.library|,
with $N = 3$ features is taken as a reference. In Fig. \ref{fig:qiskit_featuremap} the feature map circuit with 1 repetition is shown.

This circuit is parameterized on each possible $\boldsymbol{x}$.

The actual $U_\Phi(\boldsymbol{x})$ is obtained with 2 repetitions of the feature map circuit. Furthermore, the circuit needed for performing the QKE is obtained by concatenating $U^\dag_\Phi(\boldsymbol{x})$ to $U_\Phi(\boldsymbol{x})$. 

By using the methods explained in Section II, a sequence is built for each $\boldsymbol{x}_i$ with Pulser, where each pulse corresponds to a gate.
For the estimation of the training kernel matrix, a total of $m^2$ sequences that represent the evolutions $\mathcal{\xi}_M(i,j)$ are obtained, where
\begin{equation*}
    \forall (i \in [m], j \in [m]), \mathcal{\xi}_M\ket{0}:(i,j) \mapsto U^\dag_\Phi(\boldsymbol{x}_{M,i})U_\Phi(\boldsymbol{x}_{M,j})\ket{0}. 
\end{equation*}

For the estimation of the test kernel matrix, instead, a total of $sm$ sequences representing $\mathcal{\xi}_S(i,j)$ are obtained, where
\begin{equation*}
    \forall (i \in [s], j \in [m]), \mathcal{\xi}_S\ket{0}:(i,j) \mapsto U^\dag_\Phi(\boldsymbol{x}_{S,i})U_\Phi(\boldsymbol{x}_{M,j})\ket{0}. 
\end{equation*}

As an example, the sequence corresponding to $\xi_M(1,1)$ is shown in Fig. \ref{fig:seq}.

The training kernel matrix entries $(K_M)_{i,j}$ are estimated by sampling $1000$ times from the Z measurement distribution out of each sequence $\xi_M(i,j)$ and taking the square of the frequency of the $0$ outcome. 
Similarly, the testing kernel matrix entries $(K_S)_{i,j}$ are estimated by sampling $1000$ times from the Z measurements distribution out of each sequence $\xi_S(i,j)$ and taking the square of the frequency of the $0$ outcome. $K_M \in [0,1]^{m,m}$ and $K_S \in [0,1]^{s,m}$ are then used by a classical computer for training and testing the SVM algorithm on the dataset, using the Scikit-Learn SVC function with default parameters and the precomputed kernel matrices.
The performance is compared with the one obtained with a radial basis function kernel as a classical counterpart.

\section{Results}

The Hadamard, Phase ($R_z$) and CX gates are obtained with pulses using the methods shown previously. Given Eq. \ref{eq:T} with $A = \Omega_{max} = 62.83\;rad\; \mu s^{-1}$ and $\theta \in (0,2\pi]$, then, growing linearly with the desired $\theta$, it holds $T \in (0, 0.238] \;\mu s$, which is the interval of the possible duration of a pulse for single-qubit gates.
For the Rydberg $2\pi$ pulse, since a conservative blockade radius equal to $10 \;\mu m$ is considered, the corresponding maximum Rabi frequency is $\Omega_{max} = 5.42\;rad\;\mu s^{-1}$ (only for the $2\pi$ pulse), corresponding to a duration $T_{2\pi} = 2.76\;\mu s$. It was seen that reducing the considered blockade radius made the CZ pulse more imprecise.
The average QKE sequence on each pair of samples lasts approximately $75 \mu s$.

\begin{figure}[htb!]
\centerline{\includegraphics[width=0.5\textwidth]{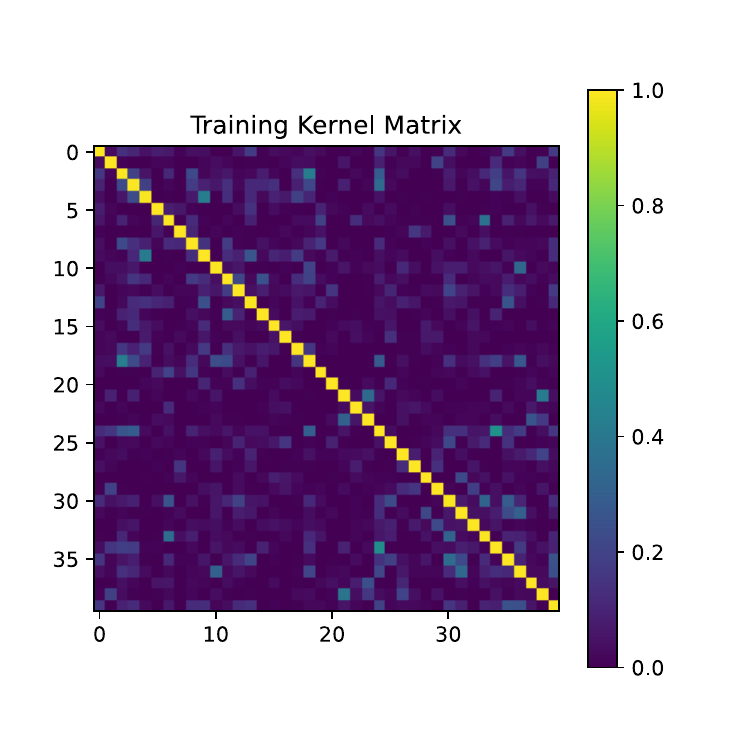}}
\caption{A heatmap of the estimated training kernel matrix, where at row $i$ and column $j$ the color corresponding to the value $K_M(i,j)$ is found. The diagonal elements are all equal to 1.}
\label{fig:training_heatmap}
\end{figure}

After running each sequence $\xi_M(i,j)$ and $\xi_S(i,j)$, the matrices $K_M$ and $K_S$ are estimated as shown in the previous sections. 
A heatmap of the estimated matrix $K_M$ is shown in Fig. \ref{fig:training_heatmap} to illustrate the values that are obtained. The closer two samples $i$ and $j$ are in the feature space, the closer the corresponding $K_M(i,j)$ is to 1. 
Naturally, $(K_M)_{i,i} = 1$ since \begin{equation*}
    \xi_M(i,i)\ket{0} = U^\dag_\Phi(\boldsymbol{x}_{M,i})U_\Phi(\boldsymbol{x}_{M,i})\ket{0} = \ket{0},
\end{equation*} and $|\braket{0|0}|^2 = 1$.
After training, all of the training samples become support vectors. The accuracy on the test set is evaluated and a mean accuracy of $75\%$ is obtained, calculated using the test kernel matrix previously estimated. For comparison with a fully classical approach, another SVC was trained using a radial basis function kernel instead of a quantum one, obtaining a mean accuracy of $65\%$, which is sensibly lower than the one obtained with the quantum approach.

\section{Discussion}
Despite the very low separation ($\Delta = 0.1$) on the dataset and the small number of samples, a high accuracy is reached. The accuracy is higher than the one obtained with the radial basis function kernel, but the main advantage is the fact that, for a high number of features, the classical approach would not only be less accurate, at least for this particular problem, but also unfeasible, whereas the quantum one does not have this critical issue. Due to the impossibility of running the sequences on a real hardware, the simulation was restricted to the case of only $N=3$ features because of the computational effort required by QuTip when the number of qubits increases. The method presented in this paper can be used for any number of features, if the chosen technology allows it, especially if spatial arrangement is permitted so that it can be fully exploited. Clearly, as the number of qubits increases, the advantage brought by the neutral atom technology is more evident, given the fact that much more 1-to-many connections among qubits are needed. 
 It is worth to notice that the purpose of quantum feature kernels makes sense for dataset such as the one used in this experiment, i.e. for data that needs to be brought in a high-dimensional feature space to be actually separable. Regarding the specific dataset used in this experiment, it is evident that it was designed to make a high classification accuracy possible using the classification method illustrated in this paper and in \cite{havlicek_supervised_2019}, and it is evident that it is just an artificial example for theoretical purposes. However, it can still be the case that in a real scenario a high-dimensional feature space could be needed, in which case quantum feature kernels present an exponential computational advantage over classical kernel computation methods. Future research should focus on the study of which real scenarios are actually of this kind, since as of today it is not clear, making quantum feature kernels only potentially advantageous in purely theoretical cases.
 On a final note, while it is true that neutral atom devices allow for a better connectivity, a pulse corresponding to a gate usually takes some microseconds, whereas on superconducting devices a gate takes a few nanoseconds. Since the duration $T$ of a pulse is inversely proportional to the maximum channel output of the device, improving the latter can drastically reduce $T$. Furthermore, a sequence such as the one used in this paper (Fig. \ref{fig:seq}) takes almost $80 \mu s$ to run, while, as of today, Pasqal devices only support sequences that last up to $10 \mu s$ due to decoherence.

\section{Conclusion}
In this paper it was shown that, using neutral atom devices, it is not only possible to work completely with the gate model, but the advantage given by the arbitrary arrangement of atoms in space can be exploited for optimizing highly entangling circuits, showing this advantage on the Quantum Kernel Estimation technique that is traditionally implemented on superconducting devices. A general method for obtaining single-qubit and multi-qubit gates was formalized, generalizing the technique proposed in \cite{silverio_pulser_2022}, and it was then used for implementing the QKE algorithm on a neutral atom device. An accuracy of $75\%$ with $3$ features was reached despite the small dimension of the dataset and the low separation $\Delta$. It was also shown how to geometrically exploit the possibility to arbitrarily arrange the atoms in space when the number of qubits is high, in order to maximally reduce the SWAP gates needed for allowing entanglement. 
This work aims to serve as a reference for future works that either wish to use the gate model on neutral atom quantum computers for any kind of problem or to further explore the possibilities that this technology allows in the field of quantum machine learning, and to provide an object of discussion and further exploration on the possible utility of quantum computing for real classification problems.

\bibliographystyle{IEEEtran}
\bibliography{conference_101719.bib}


\end{document}